\documentclass[12pt]{article}

\newcommand{\be}{\begin{equation}}
\newcommand{\ee}{\end{equation}}
\newcommand{\bi}[1]{\vspace{-3mm} \bibitem{#1}}

\usepackage{epsfig,amsmath,amssymb,graphics,graphicx} 

\begin{document}

\begin{center}
{\it International Journal of Mathematics 18 (2007) 281-299}
\end{center}

\begin{center}
{\Large \bf Fractional Derivative as Fractional Power of Derivative}
\vskip 5 mm

{\large \bf Vasily E. Tarasov }\\

\vskip 3mm

{\it Skobeltsyn Institute of Nuclear Physics, \\
Moscow State University, Moscow 119991, Russia } \\
{E-mail: tarasov@theory.sinp.msu.ru}\\
\end{center}

\begin{abstract}
Definitions of fractional derivatives as 
fractional powers of derivative operators are suggested. 
The Taylor series and Fourier series are used to define
fractional power of self-adjoint derivative operator.
The Fourier integrals and Weyl quantization procedure are applied
to derive the definition of fractional derivative operator. 
Fractional generalization of concept of stability is considered. 
\end{abstract}

\vskip 4mm

\section{Introduction}

The theory of integrals and derivatives of non-integer order goes back 
to Leibniz, Liouville, Riemann, Grunwald, and Letnikov \cite{SKM,OS}. 
Fractional analysis has found many
applications in recent studies in mechanics and physics.
The interest in fractional equations has been growing continually 
during the last few years because of numerous applications. 
In a short period of time the list of applications becomes long. 
For example, it includes the chaotic dynamics \cite{Zaslavsky1,Zaslavsky2},
material sciences \cite{Hilfer,C2,Nig1,Nig4}, 
mechanics of fractal and complex media 
\cite{Mainardi,Media,Physica2005},
quantum mechanics \cite{Laskin,Naber}, 
physical kinetics \cite{Zaslavsky1,Zaslavsky7,SZ,ZE},
plasma physics \cite{CLZ,Plasma2005}, 
electromagnetic theory \cite{Lutzen,Mil2,Plasma2005},  
astrophysics \cite{CMDA},
long-range dissipation \cite{GM,TZ2}, 
non-Hamiltonian mechanics \cite{nonHam,FracHam,FracVar},
long-range interaction \cite{Lask,TZ3,KZT}.

It is known that we can define a fractional power of 
operator \cite{Krein1,Krein2,Krein3,KZ,KS,Maslov,KM}. 
The integer power of operator can be easily realized.
Therefore, we can realize the 
fractional power as the integer power series. 
In this paper, we use equations that represent the fractional power
as a series of integer powers series.
This representation allows us to define the fractional power of operator 
as a series of integer powers of operator.
As a result, we obtain the definition of fractional derivatives
as a fractional power of derivative operator.

Note that the well-known Riemann-Liouville fractional derivative 
can be represented 
as a power series of derivatives of integer order \cite{SKM}:
\[ D^{\alpha}_{a+}=\sum^{\infty}_{n=0} A_n(x,a,\alpha) \frac{d^n}{dx^n}, \]
where
\[ A_n(x,a,\alpha)= 
\frac{(-1)^{n-1}\alpha \Gamma(n-\alpha)}{\Gamma(1-\alpha) \Gamma(n+1)}
\frac{(x-a)^{n-\alpha}}{\Gamma(n+1-\alpha)} \]
for the functions  that are analytical in the interval $(a,b)$.

In Sec. 2, we point out some well-known definitions of functions
of bounded and unbounded operators. 
In Sec. 3, the fractional derivatives are defined as 
fractional powers of coordinates that considered as Taylor series.   
In Sec. 4, the fractional derivatives are considered as 
fractional powers of coordinates that described as Fourier series
for the interval.
In Sec. 5, the fractional derivatives are defined as 
fractional powers of coordinates by using the Fourier integrals.
In Sec. 6, the fractional derivatives are defined as 
fractional powers of coordinates by using the Weyl quantization.
In Sec. 7, using fractional derivatives, we define 
the stability with respect to fractional variations.

\section{Function of bounded and unbounded operators}

Let us point out some well-known definitions of functions
of bounded and unbounded operators \cite{KZ,KS,Krein1,Krein2,Krein3}.

\subsection{Power series}

Let us consider a bounded linear operator $A$ that 
is defined on the linear space $E$, and $A\in L(E,E)$,
where $L(E,E)$ is a space of linear maps of $E$.
Suppose the function $f(x)$ is an analytical function 
of the variable $x$ such that it can be represented as a power series
\[ f(x)=\sum^{\infty}_{n=0} f_n x^n .\]
Then, we can define 
\be
f(A)=\sum^{\infty}_{n=0} f_n A^n .
\ee
The operator $f(A)$ is a linear bounded operator $A$ on space $E$.
For example, the exponential function of operator
is defined by
\[ e^A=\sum^{\infty}_{n=0} \frac{1}{n!} A^n. \]

\subsection{Cauchy's integral formula}  

The definition of operator function by power series
can be generalized for wider class of functions. 
To realize this generalization, 
we use Cauchy's integral formula instead of power series.
Cauchy's integral formula states that
\be f(z_0)=\frac{1}{2\pi i} \oint_{\Gamma} \frac{f(z)dz}{z-z_0}, \ee
where the integral is a contour integral along the contour 
$\Gamma$ enclosing the point $z_0$. 

We can define the algebraic isomorphism between 
an operator algebra and some functions \cite{Krein2}.
The function $f(z)=z$ corresponds to the operator $A$.
The function $f(z-z_0)=(z-z_0)^{-1}$ 
corresponds to the resolvent operator $R(z,A)=(zI-A)^{-1}$.
If $|z|>r_A$, where $r_A$ is a spectral radius:
\[ r_A=\lim_{n\rightarrow \infty} \sqrt[n]{||A^n||} , \]
then the resolvent operator exists. 
The function of linear bounded operator is defined by
\be
f(A)=\frac{1}{2\pi i} \oint_{\Gamma} f(z) R(z,A) dz,
\ee
where 
\[ R(z,A)=(zI-A)^{-1} , \quad z\in \rho(A) . \]
Here $\Gamma=\partial G \in \sigma(A)$, where $\sigma(A)$ 
is a spectrum of operator $A$, and $\rho(A) \subset G$. 
For example, we can define the operator
\be 
e^{At}=\sum^{\infty}_{n=0} \frac{t^n}{n!} A^n=
\frac{1}{2\pi i} \oint_{\partial G} e^{zt} R(z,A) dz 
\ee
that is corresponded to the function $\exp(zt)$, and $\sigma (A)\subset G$.

As the second example, 
the operator $E_z$ that corresponds to the Heaviside 
function $\theta (z-z_0)$, where $\theta (z-z_0)=0$ for $z_0 \ge z$, 
and $\theta (z-z_0)=1$ for $z_0 < z$ is defined by
\be
E_z=E(z,A)=
\frac{1}{2 \pi i} \oint_{\Gamma} \theta (z-z_0) R(z_0,A)dz_0 ,
\ee
and is called the spectral operator.
The operator $E_z$ can be denoted by $\theta(zI-A)$.

\subsection{Spectral representation of self-adjoint unbounded operator}

It is known that  spectral function $E_z$ exists for 
all self-adjoint operators $A$, and 
\be \label{Ax}
Ax=\int^{+\infty}_{-\infty} z dE_z x,
\ee
where
\[ ||Ax||^2=\int^{+\infty}_{-\infty} |z|^2 d(E_z x,x) < \infty . \]
Then, the function $f(A)$ of self-adjoint operator $A$ can be defined by
the equation
\be
f(A)x=\int^{+\infty}_{-\infty} f(z) dE_z x .
\ee
As operator $A$, we can consider the
self-adjoint derivative $-i \partial / \partial x$. 
For the father information about this approach, we can use Refs. 
\cite{KZ,KS,Krein3,kn1,kn2}.

\section{Fractional Derivatives by Taylor Series}

\subsection{Definition of Taylor series}

A one-dimensional Taylor series, which is an expansion of a real-valued  
function $f(x)$ about a point $x=a$, is given by
\be 
f(x)=\sum^{\infty}_{n=0} f_n (x-a)^n,
\ee
where 
\be
f_n=\frac{1}{n!} f^{(n)}(a) ,
\ee
and $f^{(n)}(a)$ is the nth derivative of $f(x)$ evaluated
at the point $x=a$.

Suppose the function $f(x)$ has all derivatives in the interval $|x-a|<a_0$,
and the condition
\[ \lim_{n\rightarrow \infty} \frac{f^{(n)}(a)}{n!} (x-a)^n=0 \]
is satisfied, then the series 
\be
f(x)=\sum^{\infty}_{n=0} \frac{f^{(n)}(a)}{n!} (x-a)^n
\ee
converges to the function $f(x)$ for all intervals $|x-a|<a'$, where $a'<a_0$.
This representation of the functions can be used to define fractional
power of operator.

\subsection{Taylor series for fractional power of coordinate}

The Taylor series for fractional power of coordinate 
$f(x)=x^{\alpha}$ about a point $x=a>0$ is 
\be \label{xa}
x^{\alpha}=\sum^{\infty}_{n=0} f_n (x-a)^n,
\ee
where 
\[ f_n=\frac{1}{n!} (x^{\alpha})^{(n)}(a)=
\frac{B(\alpha,n)}{n! a^{n-\alpha}} , \]
\[ B(\alpha,n)=\alpha (\alpha-1)(\alpha-2)...(\alpha-n+1). \]
If $m-1<\alpha<m$, then
\[ B(\alpha,n)=(-1)^{n-m} \frac{(\alpha+1-m)_{m}}{(m-\alpha)_{n-m}}=
(-1)^{n-m} \frac{\Gamma(n-\alpha) 
\Gamma(\alpha+1)}{\Gamma(m-\alpha) \Gamma(\alpha+1-m)} , \]
where
\[ (z)_m=z(z+1)...(z+m-1) . \]
For $0<\alpha<1$,
\be
B(\alpha,n)=(-1)^{n-1} \alpha (1-\alpha)_{n-1}=
(-1)^{n-1} \frac{\Gamma(n-\alpha)}{\Gamma(1-\alpha)} .
\ee 
Using
\be 
(x-a)^n=\sum^n_{k=0} (-1)^{n-k} \Bigl(^n_k\Bigr)  x^k a^{n-k}, 
\quad \Bigl(^n_k\Bigr)=\frac{n!}{(n-k)!k!} ,
\ee
we rewrite Eq. (\ref{xa}) in the form
\be \label{xa1}
x^{\alpha}=\sum^{\infty}_{n=0} \sum^n_{k=0} C(n,k,\alpha,a) x^k ,
\ee
where
\[ C(n,\alpha,a)= \frac{(-1)^{k+1} a^{\alpha-k}}{(n-k)!k!} 
\frac{\Gamma(n-\alpha)}{\Gamma(1-\alpha)} . \]

Equation (\ref{xa1}) represents the fractional power
of coordinate as a series of integer powers.
This representation allows us to define
the fractional power of operator as a
series of integer powers of operator.

It is known that self-adjoint operators have the real eigenvalues. 
Using Eq. (\ref{xa1}) with $a>0$,
we can define the fractional power of the self-adjoint operator $A$ by
\be A^{\alpha}=
\sum^{\infty}_{n=0} \frac{B(\alpha,n)}{n! a^{n-\alpha}} A^n . \ee
For the operator $A=-i {\partial}/ {\partial x}$, we have 
\be \label{Tfd}
\left(-i\frac{d}{dx}\right)^{\alpha}=
\sum^{\infty}_{n=0} \frac{B(\alpha,n)}{n! a^{n-\alpha}} 
\left(-i\frac{d}{dx}-a \right)^n,
\ee
or, in the equivalent form
\be \label{Tfd2}
\left(-i\frac{d}{dx}\right)^{\alpha}=\sum^{\infty}_{n=0} 
\sum^n_{k=0} (-i)^k C(n,k,\alpha,a) \frac{d^k}{dx^k}. 
\ee 
As a result, we get that fractional derivative
is defined as a series of integer powers of 
self-adjoint derivative operator.

\subsection{Examples of computation of fractional derivatives}

Let us consider the fractional derivative (\ref{Tfd}) of constant $c$:
\be
\left(-i\frac{d}{dx}\right)^{\alpha}c=
\sum^{\infty}_{n=0} \frac{A(\alpha,n)}{n! a^{n-\alpha}} 
\left(-i\frac{d}{dx}-a \right)^n c .
\ee
Using
\be
\left(-i\frac{d}{dx}-a \right)^n c=(-a)^n c,
\ee
we get
\be
\left(-i\frac{d}{dx}\right)^{\alpha}c=
a^{\alpha} c \sum^{\infty}_{n=0} (-1)^n \frac{B(\alpha,n)}{n!} .
\ee
If $0<\alpha<1$, then 
\be
\left(-i\frac{d}{dx}\right)^{\alpha}c=
-a^{\alpha} c \sum^{\infty}_{n=0} 
\frac{\Gamma(n-\alpha)}{\Gamma(n+1) \Gamma(1-\alpha)} .
\ee 

Let us consider the fractional derivative of a power $x^m$.
From (\ref{Tfd2}), 
\be 
\left(-i\frac{d}{dx}\right)^{\alpha}x^m=\sum^{\infty}_{n=0} 
\sum^n_{k=0} (-i)^k C(n,k,\alpha,a) (x^m)^{(k)}. 
\ee 
Using
\be
(x^m)^{(k)}=m(m-1)...(m-k+1) x^{m-k}
=\frac{m!}{(m-k)!}  x^{m-k}
\ee
for $k \le m$, and $(x^m)^{(k)}=0$ for $k>m$, we obtain 
\be
\left(-i\frac{d}{dx}\right)^{\alpha} x^m=\sum^{m}_{n=0} 
\sum^n_{k=0} (-i)^k C(n,k,\alpha,a) \frac{m!}{(m-k)!} x^{m-k}. 
\ee

\section{Fractional Derivatives by Fourier Series}

\subsection{Fourier series}

Fourier series of a function $f(x) \in L_2[-l,l]$ is an 
expansion in terms of an infinite sum of sines and cosines.
Since sines and cosines form a complete orthogonal 
system over $[-l,l]$, the Fourier series is given by
\be 
f(x)=\frac{a_0}{2}+\sum^{\infty}_{n=1}
\left[ a_n \cos \left(\frac{\pi nx}{l} \right)+
b_n \sin \left(\frac{\pi nx}{l}\right) \right], 
\ee
where
\[ a_n = \frac{1}{l} \int^{+l}_{-l} f(x) \cos \left(\frac{\pi nx}{l} \right)dx , 
\quad
b_n = \frac{1}{l} \int^{+l}_{-l} f(x) \sin \left(\frac{\pi nx}{l} \right)dx , \]
and $n$ is a positive integer number. 

Let us consider the Fourier series for $f(x)=|x|^{\alpha} \in L^2 [-l;l]$ ,
where $\alpha$ is a positive fractional power.
The Fourier series of this function for $x \in [-l,+l]$ is 
\be \label{xa2}
|x|^{\alpha}=\frac{a_0}{2}+\sum^{\infty}_{k=1} a_k 
\cos \left( \frac{\pi k x}{l} \right) , 
\ee
where 
\[ a_k=\frac{1}{l} 
\int^{l}_{-l} |y|^{\alpha} \cos \left(\frac{\pi ky}{l}\right)dy
=\frac{2}{l} 
\int^{l}_{0} y^{\alpha} \cos \left(\frac{\pi ky}{l} \right)dy , 
\quad k=0,1,2,.... \]
The cosine can be represented as a power series.
Therefore, equation (\ref{xa2}) allows us to present 
the fractional power as a series of integer powers series.
Then, we can define fractional derivative on the interval $[-l,l]$
as a fractional power of derivative.

\subsection{Fractional derivative}

The fractional power of derivative
operator for the interval $[-l;l]$, we can define by
\be  \label{ssFD}
\left(-i\frac{d}{dx}\right)^{\alpha}=
\frac{a_0}{2}+\sum^{\infty}_{k=1} a_k 
\cos \left(-i \frac{\pi k }{l} \frac{d}{dx} \right) .
\ee
It is known that we can define $\exp(A)$ for 
the self-adjoint operator $A=-id/dx$ by 
\be \label{C1}
e^{-id/dx}=\sum^{\infty}_{n=0}\frac{1}{n!}A^n=\sum^{\infty}_{n=0} 
\frac{(-i)^n}{n!} 
\left(\frac{d}{dx} \right)^n .
\ee
Using
\[ \cos(A)=\frac{1}{2} \left( e^{iA}+e^{-iA} \right)=
\frac{1}{2} \left( \sum^{\infty}_{n=0}\frac{(iA)^n}{n!} 
+ \sum^{\infty}_{n=0}\frac{(-iA)^n}{n!} \right) = \]
\[ =\sum^{\infty}_{n=0} \frac{1}{2 (n!)} \left[ \left( A \right)^n 
+ \left( - A \right)^n  \right] =
\sum^{\infty}_{n=0}\frac{1+(-1)^n}{2 (n!)} A^n =
\sum^{\infty}_{m=0} \frac{1}{(2m)!} A^{2m} , \]
we have
\be \label{cos}
\cos \left(-i\frac{d}{dx}\right)=
\sum^{\infty}_{m=0} \frac{1}{(2m)!} \left(\frac{d}{dx} \right)^{2m} .
\ee
Equation (\ref{C2}) and (\ref{cos}) allows us 
to define the fractional power of operator as a
series of integer powers of operator.
Using (\ref{cos}), we rewrite (\ref{ssFD}) as
\be
\left(-i \frac{d}{dx} \right)^{\alpha}=
\frac{a_0}{2}+\sum^{\infty}_{k=1} a_k 
\sum^{\infty}_{m=0} \frac{1}{(2m)!} 
\left( \frac{\pi k}{l} \right)^{2m} \left(\frac{d}{dx} \right)^{2m} , 
\ee
or by the equivalent equation
\be
\left(-i\frac{d}{dx} \right)^{\alpha}=\frac{a_0}{2}+\sum^{\infty}_{m=0} S_m 
\left(\frac{d}{dx} \right)^{2m} , 
\ee
where
\be
S_m =\frac{1}{(2m)!} \sum^{\infty}_{k=1} 
\left( \frac{\pi k}{l} \right)^{2m} a_k .
\ee

Let us compute the coefficients $a_k$  by 
\[ a_k=\frac{1}{l^{\alpha}} \int^l_0 y^{\alpha} \cos(\pi k y) dy. \]
As a result, we obtain
\[ a_k= 2^{\alpha} \pi^{-1/2-\alpha} l^{-\alpha} k^{-1-\alpha}
\Biggl( \frac{ 2^{-\alpha} \pi^{-1/2+\alpha} k^{\alpha} 
(\alpha+3) \sin(\pi k)}{ (\alpha+1)(\alpha+3)}+ \]
\[ +\frac{2^{-\alpha} \pi^{-1/2+\alpha} k^{\alpha} 
[\pi k \cos(\pi k)-\sin(\pi k)]}{\alpha+1}+
\frac{2^{-\alpha} \sqrt{k} \alpha L(\alpha+1/2,3/2,\pi k) 
\sin(\pi k)}{\alpha+1}- \]
\[ -\frac{2^{-\alpha} [\pi k \cos(\pi k)-\sin(\pi k)] 
L(\alpha+3/2,1/2,\pi k) }{\sqrt{k} \pi (\alpha+1)} \Biggr) , \]
where $L(\mu,\nu,z)$ is the Lommel function \cite{Luke}.

\subsection{Complex Fourier series}

The real-valued function $f(x)$, which is defined on [-L/2,L/2],
can be presented by
\be f(x) = \sum_{n=-\infty}^{\infty} f_n e^{i(2\pi n/L)x} , \ee
where 
\be \label{fn}
f_n = \frac{1}{L} \int_{-L/2}^{+L/2} f(x) e^{-i(2\pi nx/L)} dx. 
\ee

The operator function $f(A)$ is defined by
\be \label{fA}
f(A)=\sum_{n=-\infty}^{\infty} f_n e^{i (2 \pi n/L)A} , \ee
where $f_n$ are defined in (\ref{fn}), and
\be \label{expp}
e^{i(2\pi n/L)A} =\sum^{\infty}_{k=0} i^k \frac{(2 \pi n/L)^k}{k!} A^k.
\ee
Substitution of (\ref{expp}) into (\ref{fA}) gives
\be f(A)=\sum_{n=-\infty}^{\infty} \sum^{\infty}_{k=0} 
f_n  \frac{(i 2 \pi n/L)^k}{k!} A^k . \ee
Then the fractional power of self-adjoint derivative on $[-L/2,+L/2]$
can be presented by
\be \label{CFS}
\left(- i \frac{d}{dx}\right)^{\alpha}=
\sum_{n=-\infty}^{\infty} \sum^{\infty}_{k=0} 
f(n,k)  \left(\frac{d}{dx}\right)^k ,
\ee
where 
\be f(n,k)=\frac{(2 \pi n)^k}{k! L^{k+1}} 
\int_{+L/2}^{-L/2} |x|^{\alpha} \cos(nx) dx . \ee
Equation (\ref{CFS}) represents the fractional power
as a series of integer powers.
This representation defines
the fractional power of derivative  as a
series with integer powers of derivatives.

\section{Fourier Transform and Fourier Integral}

\subsection{Fractional derivative by Fourier integral}

Let us consider a function $f(x)$ with $n$ variables $x$. 
Suppose $A_1$, $A_2$,...,$A_n$ are $n$ elements of commutative 
operator algebra ${\cal A}$.
For example, ${\cal A}$ is an algebra of operators in linear space. 
We denote by $\tilde f$ the Fourier transform for $f(x)$:
\be \label{W1}
\tilde f(y)=\int^{+\infty}_{-\infty}f(x) e^{-ixy} dx ,
\quad xy=x_1y_1+...+x_ny_n ,  
\ee
where
\be 
f(x) =\frac{1}{(2\pi)^n} \int^{+\infty}_{-\infty}
\tilde f(y) e^{i(y_1x_1+...+y_nx_n)}dy=
\frac{1}{(2\pi)^n} \int^{+\infty}_{-\infty}dy \tilde f(y) e^{iyA} .  \ee
The operator function $f(A)$ of elements $A_1,...,A_n$ 
is defined by 
\be \label{W2}
f(A) =\frac{1}{(2\pi)^n} \int^{+\infty}_{-\infty}\tilde f(y) e^{i(y_1A_1+...+y_nA_n)}dy=
\frac{1}{(2\pi)^n} \int^{+\infty}_{-\infty}dy \tilde f(y) e^{iyA} .  \ee
Substitution of Eq. (\ref{W1}) into Eq. (\ref{W2}) gives
\be \label{W3}
f(A) =\frac{1}{(2\pi)^n}  \int^{+\infty}_{-\infty}\tilde f(y) e^{i y A}dy=
\frac{1}{(2\pi)^n} \int^{+\infty}_{-\infty} dy \int^{+\infty}_{-\infty} dx f(x) e^{iy(A-x)} .  \ee
In general, we must have the 
exact definition of these integrals and description of possible 
class of symbols and algebras of elements $A_1,..,A_n$.

To define the fractional power of the operator $A$, 
we use (\ref{W2}) in the form
\be \label{W4} 
A^{\alpha} =
\frac{1}{(2\pi)^n} \int^{+\infty}_{-\infty} dx \int^{+\infty}_{-\infty} dy |x|^{\alpha} e^{iy(A-x)} .  \ee
This equation can be considered as a definition of fractional power
of operator A. 

For $f(x)=|x|^{\alpha}$, where $\alpha \not=-1,-3,...$, 
we have \cite{BP}:
\[ \tilde f(y)= \int^{+\infty}_{-\infty} dx |x|^{\alpha}e^{-ixy} =
-2 \sin(\pi \alpha/2) \Gamma(\alpha+1)  
|y|^{-\alpha-1} . \]
Then Eq. (\ref{W4}) gives
\be \label{A1}
A^{\alpha} =
-\frac{2\sin(\pi \alpha/2) \Gamma(\alpha+1)}{(2\pi)^n}
\int^{+\infty}_{-\infty} dy |y|^{-\alpha-1} e^{iyA} . 
\ee
For the self-adjoint derivative operators 
\[ A_1=-i\frac{\partial}{\partial x_1},..,
A_n=-i\frac{\partial}{\partial x_n}, \]
equation (\ref{A1}) is
\be 
\left( -i\frac{\partial}{\partial x} \right)^{\alpha} =
-\frac{2\sin(\pi \alpha/2) \Gamma(\alpha+1)}{(2\pi)^n}
\int^{+\infty}_{-\infty} dy |y|^{-\alpha-1} 
\exp \left( y \frac{\partial}{\partial x} \right) . 
\ee
As a result, we obtain the fractional derivative operator 
as a fractional power of self-adjoint derivative operator.

\subsection{Fractional power of self-adjoint derivative operator}

Let us consider the self-adjoint derivative operators
\be \label{A2}
D_x= -i\frac{\partial}{\partial x}=\
\left( -i\frac{\partial}{\partial x_1},...,-
i\frac{\partial}{\partial x_n}\right). \ee
It is easy to prove that $f(p)=p^{\alpha} \in S^{\infty}(\mathbb{R}^n_p)$. 
Here, $S^{\infty}$ is the space of symbols  that are slowly 
growth on the infinity. This space is defined as
\[ S^{\infty} (\mathbb{R}^n) =\cup_l \cap_k S^k_l (\mathbb{R}^n) ,  \]
where $S^k_l (\mathbb{R}^n) $ is a space of functions from the 
class $C^k(\mathbb{R}^n)$ with the norm
\[  ||f||_{S^k_l (\mathbb{R}^n) }=\sup_{\mathbb{R}^n}(1+|x|^2)^{l/2} 
\Bigl( \sum_{|a|} ||f^{(a)} (x)||  \Bigr) . \]
The space $S^{\infty}(\mathbb{R}^n)$ is defined by
\[ \exists r, \quad \forall s \quad
||f||_{r,s}=\sup_x \{ \; (1+|x|)^r \left| \left( 
\frac{\partial}{\partial x}\right)^s f(x) \right| \; \} <\infty . \]

The fractional powers of the operator (\ref{A2})
are elements of algebra ${\cal L}(S^{\infty},S^{\infty})$ 
of all continuous linear maps of the space $S^{\infty}(\mathbb{R}^n) $.
Then, the fractional derivative operator
\be \label{2-2n} 
D^{\alpha}_{x}=\left(-i\frac{\partial}{\partial x}\right)^{\alpha}  \ee
is a fractional power of self-adjoint derivative operator.
The operator (\ref{2-2n}) acts on the arbitrary 
function $u(x) \in C^{\infty}(\mathbb{R}^n)$ by
\be \label{2-3n} 
\left(-i\frac{\partial}{\partial x}\right)^{\alpha}u(x)= 
\tilde F_{p\rightarrow x} p^{\alpha} F_{y \rightarrow p} u(y), \ee
where
\be \label{2-4n} F_{y\rightarrow p}u (y)=
\left(\frac{1}{2\pi i}\right)^{n/2} 
\int^{+\infty}_{-\infty} e^{ipy} u(y) dy,  \ee
is the direct Fourier transform, and 
\be \label{2-5n} \tilde F_{p \rightarrow x}\Psi (p)=
\left(\frac{1}{2\pi i}\right)^{n/2} 
\int^{+\infty}_{-\infty} e^{-ipx} \Psi(p) dp  \ee
is the Fourier transform.\\

{\bf Proposition.} {\it The operator 
\be \label{2-7n} 
D^{\alpha}_x=\left(-i\frac{\partial}{\partial x}\right)^{\alpha} \ee
has the symbol}
\be \label{2-8n} 
symb \{D^{\alpha}_x\}(p) =p^{\alpha} . \ee

It is not hard to prove this proposition. 
Using Eq. (\ref{2-3n}), we get
\[ D^{\alpha}_x e_p(x)=
\left(-i\frac{\partial}{\partial x}\right)^{\alpha}e_p(x)=
\left(-i\frac{\partial}{\partial x}\right)^{\alpha} e^{ipx}=
\tilde F_{z\rightarrow x} z^{\alpha} F_{y\rightarrow z} e^{ipy}= \]
\[ =\tilde F_{z\rightarrow x} \left( z^{\alpha} (2\pi i)^{n/2} 
\delta(z-p) \right) = \]
\[ =\tilde F_{z\rightarrow x} \left( p^{\alpha} (2\pi i)^{n/2}
 \delta(z-p) \right) = p^{\alpha} e^{ipx}=p^{\alpha} e_p(x) . \]
As a result, we obtain
\be \label{ddd} D^{\alpha}_x e_p(x)
=p^{\alpha} e_p(x) . \ee
Multiplying both sides of (\ref{ddd}) on $e_{-p}(x)$, 
we get $e_{-p}(x) D^{\alpha}_x e_p(x)=p^{\alpha}$
that proves (\ref{2-8n}).

\section{Fractional Derivative by Quantization Map}

\subsection{Quantization procedure for coordinate representation}

Let us consider the quantum mechanics in coordinate representation.
It is known that quantization $\hat Q$ is a linear map of coordinate $q$ 
and momentum $p$ into self-adjoint operators 
\be 
\hat Q(q)=\hat q=q , \quad \hat 
Q(p)= \hat p =-i \hbar \frac{\partial}{\partial q}, 
\quad \hat Q(1)=\hat I .
\ee
Using linearity of quantization map \cite{BerShub,Ber}, we get
\[ \hat Q(aq+bp)=a\hat q +b \hat p . \]
Obviously, we have
\[ \hat Q([aq+bp]^n)=[a\hat q +b \hat p]^n . \]
Using the power series 
\be
\exp(x)=\sum^{\infty}_{n=0} \frac{1}{n!} x^n ,
\ee
we get 
\be
\hat Q \Bigl( \exp[(i/\hbar)(aq+bp)] \Bigr)= 
\exp[(i/\hbar)(a\hat q+b \hat p)] .
\ee
This allows us to define the function of operators $\hat q$ and  $\hat p$
by using the Fourier transforms \cite{BerShub,Ber,BJ,And,Quant,kn1,kn2}.

\subsection{Weyl quantization}

Canonical quantization defines a map
of real-valued functions into self-adjoint operators 
\cite{BerShub,Ber,BJ,And,Quant,kn1,kn2}.
A classical observable is described by some real-valued function
$A(q,p)$ from a function space ${\cal M}$.
Quantization of this function leads to self-adjoint
operator $\hat A(\hat q, \hat p)$ from some operator space $\hat {\cal M}$.

Let us consider main points of the usual method of canonical
quantization \cite{BerShub,BJ,kn1,kn2}.
Suppose $q_k$ are canonical coordinates and $p_k$
are canonical momenta, where $k=1,...,n$.
The basis of the space ${\cal M}$ of
functions $A(q,p)$ is defined by functions
\be \label{f1}
W(a,b,q,p)=e^{(i/\hbar)(aq+bp)} \ ,
\quad aq=\sum^{n}_{k=1} a_k q_k  \ .
\ee
Quantization transforms coordinates $q_k$ and momenta $p_k$ to
operators $\hat q_k$ and $\hat p_k$. Weyl quantization of the basis
functions (\ref{f1}) leads to the Weyl operators
\be \label{f2} 
\hat Q \Bigl( W(a,b,q,p) \Bigr)=
\hat W(a,b,\hat q, \hat p)=
e^{(i/\hbar)(a\hat q+b \hat p)} \ ,
\quad a \hat q=\sum^{n}_{k=1} a_k \hat q_k \ .
\ee
Operators (\ref{f2}) form a basis of the operator space $\hat {\cal M}$.
Classical observable, which is characterized by the function $A(q,p)$,
can be represented in the form
\be \label{f3}
A(q,p)=\frac{1}{(2\pi \hbar)^n} 
\int^{+\infty}_{-\infty} \tilde A(a,b) W(a,b,q,p) d^n a d^n b \ ,
\ee
where 
\be \label{tildeA}
\tilde A(a,b) =
\frac{1}{(2\pi \hbar)^n} \int^{+\infty}_{-\infty} 
\tilde A(a,b) W(a,b,q,p) d^n a d^n b \ ,
\ee
i.e., $\tilde A(a,b)$ is the Fourier image of the function $A(q,p)$.
Quantum observable $\hat A(\hat q,\hat p)$, 
which corresponds to $A(q,p)$, is 
\be \label{f4} 
\hat Q \Bigl( A(q,p) \Bigr)=\hat A(\hat q, \hat p)=\frac{1}{(2\pi \hbar)^n}
\int^{+\infty}_{-\infty} \tilde A(a,b) \hat W(a,b,\hat q,\hat p) d^na d^nb  \ .
\ee
This formula can be considered as an operator expansion for
$\hat A(\hat q, \hat p)$ in the operator basis (\ref{f2}).
Substitution of (\ref{tildeA}) into (\ref{f4}) gives
\be \label{f41} 
\hat A(\hat q,\hat p)=\frac{1}{(2\pi \hbar)^{2n}} 
\int^{+\infty}_{-\infty} d^na \; d^nb  \int^{+\infty}_{-\infty} 
d^nq \; d^np \; A(q,p) \hat W(a,b, \hat q - q \hat I,\hat p - p \hat I) . \ee
The function $A(q,p)$ is called the Weyl symbol of the
operator $\hat A (\hat q, \hat p)$.
Canonical quantization defined by (\ref{f41}) is called the Weyl
quantization. The Weyl operator (\ref{f2}) in formula (\ref{f41}) 
leads to the Weyl quantization. 
Another basis operator leads to different quantization scheme \cite{BJ}.

\subsection{Fractional derivative by Weyl quantization map}

Let us consider a quantization map of real-valued function $f(p)=|p|^{\alpha}$
into self-adjoint operator.
Quantization of this function leads to some self-adjoint
operator $\hat f(\hat p)$ from the operator space $\hat {\cal M}_p$,
where $\hat p_k=-i\partial/\partial x_k$ and $k=1,...,n$. Then 
\[ D^{\alpha}_x=f(\hat p)=\hat p^{\alpha}=
(-i\partial/\partial x_k)^{\alpha}. \]
The basis of the space ${\cal M}_p$ is defined 
by functions
\be \label{f1n}
W(a, p)=e^{i bp} \ , \quad ap=\sum^{n}_{k=1} a_k p_k  .
\ee
Quantization maps $p_k$ into $\hat p_k=-i \partial / \partial x_k$. 
Weyl quantization of the functions (\ref{f1n}) leads to 
\be \label{f2n}
\hat Q \Bigl( W(a,b)\Bigr) =
\hat W(a,\hat p)=e^{i a \hat p} ,
\quad  a \hat p=\sum^{n}_{k=1} a_k \hat p_k  .
\ee
The operators (\ref{f2n}) form a basis of 
the operator space $\hat {\cal M}_p$.
Using Fourier transform, the function $f(p)=|p|^{\alpha}$ 
can be presented by
\be \label{f3n}
f(p)=\frac{1}{(2\pi)^{n/2}} \int^{+\infty}_{-\infty} \tilde f(a) W(a,p) \, d^n a  ,
\ee
where 
\be \label{C2}
\tilde f(a)=\frac{1}{(2\pi)^{n/2}} \int^{+\infty}_{-\infty} f(p) W(a,p) \, d^n p, 
\ee 
i.e., $\tilde f(a)$ is the Fourier image of the function $f(p)$.
Quantum observable $\hat f(\hat p)$, which corresponds to $f(p)$, 
is defined by the formula
\be \label{f4n} 
\hat Q \Bigl( f(p) \Bigr)= \hat f(\hat p)=\frac{1}{(2\pi)^{n/2}}
\int^{+\infty}_{-\infty} \tilde f(a) \hat W(a,\hat p) \, d^n a   .
\ee
This formula can be considered as an operator expansion for
$\hat f(\hat p)$ in the operator basis (\ref{f2}).
From (\ref{C2}) and (\ref{f4n}), we obtain 
\be \label{f41n} 
\hat f(\hat p)=\frac{1}{(2\pi)^{n}} \int^{+\infty}_{-\infty} d^na
\int^{+\infty}_{-\infty} d^np \;  f(p) 
\hat W(a,\hat p - p \hat I) . \ee
The function $f(p)$ is the Weyl symbol of the
operator $\hat f (\hat p)$.

For $f(p)=|p|^{\alpha}$, 
where $\alpha$ is a positive real number, we obtain
\be 
D^{\alpha}_x=\left(-i \frac{d}{dx}\right)^{\alpha} =
\frac{1}{(2\pi)^{n}} \int^{+\infty}_{-\infty} d^na 
\int^{+\infty}_{-\infty} d^np  \; |p|^{\alpha}
\hat W(a,\hat p - p \hat I) . \ee
As a result, we have the definition of fractional 
derivatives on $\mathbb{R}^n$ as a fractional power of self-adjoint derivative.


\section{Fractional Stability}

In this section, we use the fractional generalization of 
variations of variables. 
Fractional integrals and derivatives are used for
stability problems \cite{S1,S2,S3,S4,S5}. 
In this paper, we consider the properties of dynamical systems
with respect to fractional variations \cite{FracVar}.
We formulate stability with respect to
motion changes at fractional changes of variables.
Some systems can be unstable "in sense of Lyapunov", 
and be stable with respect to fractional variations.

\subsection{Fractional variation derivative}

Let us consider dynamical system that is defined by the 
ordinary differential equations. 
Suppose that the motion of dynamical system is described by 
the equations
\be \label{1}
\frac{d}{dt} y_k=F_k(y) ,\quad k=1,...,n.
\ee
Here $y_1,...,y_n$ be real variables that define
the state of dynamical system.

Let us consider the variation $\delta y_k$ of variables $y_k$.
The unperturbed motion is satisfied to 
zero value of variation $\delta y_k=0$.
The variation describes that as function $f(y)$ changes 
at changes of argument $y$.
The first variation describes changes of function
with respect to the first power of changes of $y$:
\be
\delta f(y) =D^{1}_y f(y) dy,
\ee
where 
\[ D^{1}_y f(y)=\frac{\partial f(y)}{\partial y} . \]
The second variation describes changes of function
with respect to the second power of changes of $y$:
\be
\delta^2 f(y) =D^{2}_y f(y) (dy)^2.
\ee
The variation $\delta^{n}$ of integer order $n$ is defined by the derivative
of integer order $D^{n}_y f(y)=\partial^n f /\partial y^n$.

Let us define the variation of fractional order as a fractional 
exterior derivative of the function (zero-form) by the equation
\be \label{fv}
\delta^{\alpha} f=D^{\alpha}_{y} f \; (\delta y)^{\alpha} ,
\ee
where $D^{\alpha}_{y}$ is a fractional derivative with respect to $y$. 

The fractional variation of order $\alpha$ describes the function 
$f(y)$ changes with respect to fractional power of variable $y$ changes.
The variation of fractional order is defined by the derivative
of fractional order.

\subsection{Equations for fractional variations}

Let us derive the equations for fractional variations 
$\delta^{\alpha} y_k$.
We consider the fractional variation of equation (\ref{1}) in the form:
\be \label{fvar1}
\delta^{\alpha}  \frac{d}{dt} y_k=\delta^{\alpha}  F_k(y) ,\quad k=1,...,n.
\ee
Using the definition of fractional variation (\ref{fv}), we have 
\be \label{Sta1}
\delta^{\alpha}  F_k(y)=
[D^{\alpha}_{y_l} F_k] (\delta y_l)^{\alpha}  ,\quad k=1,...,n.
\ee
From Eq. (\ref{Sta1}), and the property of variation 
\be
\delta^{\alpha} \frac{d}{dt}y_k =\frac{d}{dt} \delta^{\alpha} y_k, 
\ee
where $y_k=y_k(t,a)$, we obtain
\be \label{fvar3}
\frac{d}{dt} \delta^{\alpha} y_k= 
[D^{\alpha}_{y_l} F_k] \; (\delta y_l)^{\alpha} ,\quad k=1,...,n.
\ee
Note that in the left hand side of Eq. (\ref{fvar3}),
we have fractional variation of $\delta^{\alpha} y_k$,
and in the right hand side - 
fractional power of variation $(\delta y_k)^{\alpha}$.

Let us consider the fractional variation of the variable $y_k$.
Using Eq. (\ref{fv}), we get
\be \label{yy}
\delta^{\alpha}  y_k=
[D^{\alpha}_{y_l} y_k] \; (\delta y_l)^{\alpha}  ,\quad k=1,...,n.
\ee
For the Riemann-Liouville fractional derivative, 
\[ D^{\alpha}_{y_l} y_k \not=0 \] 
if $k \not=l$.
Therefore for simplification of our transformations, 
we use the fractional derivative as a fractional power of derivative 
\be \label{25}
D^{\alpha}_{y_l} y_k =
\delta_{kl} D^{\alpha}_{y_l} y_l  ,
\ee
where $\delta_{kl}$ is a Kronecker symbol.
Substituting Eq. (\ref{25}) into Eq. (\ref{yy}), we can
express the fractional power of variations $(\delta y_k)^{\alpha}$
through the fractional variation $\delta^{\alpha} y_k$:
\be \label{yy2}
(\delta y_k)^{\alpha}=D^{\alpha}_{y_k} y_k  \delta^{\alpha} y_k. 
\ee
Substitution of Eq. (\ref{yy2}) into Eq. (\ref{fvar3}) gives 
\be \label{fvar2}
\frac{d}{dt} \delta^{\alpha} y_k=
\left[ D^{\alpha}_{y_l} y_l  \right] \left[ D^{\alpha}_{y_l}F_k \right] 
\delta^{\alpha} y_l .
\ee
Here we mean the sum on the repeated index $l$ from 1 to $n$.
Equation (\ref{fvar2}) is equations for fractional variations.
Let us denote $x_k$ the fractional variations $\delta^{\alpha} y_k$:
\be
x_k=\delta^{\alpha} y_k=
\left[ D^{\alpha}_{y_k} y_k \right] (\delta y_k)^{\alpha} .
\ee

As a result, we obtain the differential equation
for fractional variations
\be \label{fxax2}
\frac{d}{dt} x_k=a_{kl}(\alpha) x_l,
\ee
where 
\be 
a_{kl}(\alpha)= \left[ D^{\alpha}_{y_l} y_l  \right]
D^{\alpha}_{y_l}F_k . \ee
Using the matrix $X^t=(x_1,...,x_n)$, and $A_{\alpha}=||a_{kl}(\alpha)||$, 
we can rewrite Eq. (\ref{fxax2})
in the matrix form
\be \label{Sta2}
\frac{d}{dt} X=A_{\alpha}X . \ee
Equation (\ref{Sta2}) is a linear differential equation.
To define the stability with respect to fractional variations, 
we consider the characteristic equation 
\be
Det[A_{\alpha}-\lambda E]=0 
\ee
with respect to $\lambda$.
If the real part $Re[\lambda_k]$ of all eigenvalues $\lambda$ 
for the matrix $A_{\alpha}$ are negative, then the 
unperturbed motion is asymptotically stable with 
respect to fractional variations. 
If the real part $Re[\lambda_k]$ of one of the eigenvalues 
$\lambda$ of the matrix $A_{\alpha}$ is positive, 
then the unperturbed motion is unstable
with respect to fractional variations. 

A system is said to be stable with respect to fractional variations
if for every $\epsilon$, there is a $\delta_0$ such that:
\be  \|\delta^{\alpha} y(t_0)\| < \delta_0 \quad => 
\quad \| \delta^{\alpha} y(\alpha,t)\| < \epsilon 
\quad \forall t \in \mathbb{R}_{+} . \ee
The dynamical system is said to be asymptotically stable 
with respect to fractional variations 
$\delta^{\alpha} y(t,\alpha)$ if as
\be
t \rightarrow \infty, \quad 
\|\delta^{\alpha} y(t,\alpha)\| \rightarrow 0 . 
\ee

The concept of stability with respect to fractional variations
is wider than the usual Lyapunov or asymptotic stability. 
Fractional stability includes concept of "integer" stability 
as a special case $\alpha=1$.
Some systems can be unstable with respect to first variation
of states, and be stable with respect to fractional variation.
Therefore fractional derivatives expand our possibility to research
the properties of dynamical systems.


\end{document}